\documentclass[aps,a4paper,twocolumn,superscriptaddress,english,longbibliography]{revtex4-1}
\usepackage{amsmath,amssymb}
\usepackage{graphicx,epsfig,sidecap,wasysym,soul}
\usepackage{color,epstopdf}
\definecolor{blue}{rgb}{0.4, 0.5, 0.9}
\definecolor{red}{rgb}{1, 0.13, 0.32}
\definecolor{green}{rgb}{0, 0.5, 0}
\usepackage[colorlinks,bookmarks=false,citecolor=green,linkcolor=red,urlcolor=blue]{hyperref}

\def\be{\begin{equation}}
\def\ee{\end{equation}}
\def\bea{\begin{eqnarray}}
\def\eea{\end{eqnarray}}

\begin{document}
\title{Chain breaking and Kosterlitz-Thouless scaling at the many-body localization transition in the random field Heisenberg spin chain}
\author{Nicolas Laflorencie}
\email{nicolas.laflorencie@irsamc.ups-tlse.fr}
\affiliation{Laboratoire de Physique Th\'eorique, IRSAMC, Universit\'e de Toulouse, CNRS, UPS, France}
\author{Gabriel Lemari\'e}
\email{lemarie@irsamc.ups-tlse.fr}
\affiliation{Laboratoire de Physique Th\'eorique, IRSAMC, Universit\'e de Toulouse, CNRS, UPS, France}
\affiliation{MajuLab, CNRS-UCA-SU-NUS-NTU International Joint Research Unit, Singapore}
\affiliation{Centre for Quantum Technologies, National University of Singapore, Singapore}
\author{Nicolas Mac\'e}
\email{nicolas.mace@irsamc.ups-tlse.fr}
\affiliation{Laboratoire de Physique Th\'eorique, IRSAMC, Universit\'e de Toulouse, CNRS, UPS, France}
\begin{abstract}
Despite tremendous theoretical efforts to understand subtleties of the many-body localization (MBL) transition, many questions remain open, in particular concerning its critical properties.
Here we make the key observation that MBL in one dimension is accompanied by a spin freezing mechanism which {causes chain breakings} in the thermodynamic limit.
Using analytical and numerical approaches, we show that such chain breakings directly probe the typical {localization length}, and {that their} scaling properties {at} the MBL transition agree with the Kosterlitz-Thouless scenario predicted by phenomenological renormalization group approaches. 
\end{abstract}
\date{\today}
\maketitle

{\noindent\bf{{Introduction---}}}
The field of interacting quantum systems in the presence of disorder has attracted a lot of attention over the last two decades. Besides tremendous theoretical efforts~\cite{jacquod_emergence_1997,gornyi_interacting_2005,basko_metal-insulator_2006,znidaric_many-body_2008,pal_many-body_2010,bardarson_unbounded_2012,nandkishore_many-body_2015,imbrie_many-body_2016,abanin_recent_2017,alet_many-body_2018,abanin_many-body_2019}, a growing number of experimental realizations have also emerged, either based on ultracold atoms or trapped ions~\cite{schreiber_observation_2015,choi_exploring_2016,smith_many-body_2016,kaufman_quantum_2016,lukin_probing_2019}, superconducting qubits~\cite{roushan_spectroscopic_2017,xu_emulating_2018,chiaro_growth_2019}, solid-state setups~\cite{ovadia_evidence_2015,de_luca_dynamic_2015,wei_exploring_2018}. Building on this collective movement (for recent reviews, see Refs.~\cite{nandkishore_many-body_2015,imbrie_many-body_2016,abanin_recent_2017,alet_many-body_2018,abanin_many-body_2019}), it is nowadays well-admitted that upon increasing disorder several low-dimensional quantum interacting systems can undergo a transition towards a many-body localized (MBL) phase. This non-ergodic regime is reasonably well-characterized, mostly by exact diagonalization (ED) techniques~\cite{luitz_many-body_2015,pietracaprina_shift-invert_2018} showing various properties of MBL states: Poisson spectral statistics, area-law entanglement, emerging integrability, logarithmic spreading of entanglement, eigenstates multifractality. While typically limited to $L\sim 20$ interacting 2-level systems (due to exponentially small  level-spacings $\propto 2^{-L}$ at high energy), ED studies have nevertheless showed a clear ergodicity breaking transition for the random-field spin-$1/2$ Heisenberg chain model
\be
{\cal H}=\sum_{i=1}^{L}\left(\vec S_i\cdot\vec S_{i+1}-h_iS_i^z\right),
\label{eq:H}
\ee
where $h_i$ are independently drawn form a uniform distribution $[-h,h]$. Despite recent debates~\cite{devakul_early_2015,doggen_many-body_2018,panda_can_2020,suntajs_quantum_2019,chanda_time_2020,abanin_distinguishing_2019,sierant_thouless_2019}, $h_c\sim 4$ is the most accepted numerical estimate~\cite{pal_many-body_2010,luitz_many-body_2015} for the  critical disorder strength at infinite temperature. 

A more serious issue concerns the universality class of the ergodic-MBL transition, for which numerical simulations yield a correlation length $\xi\sim |h-h_c|^{-\nu}$ with an exponent $\nu\approx1$ for both sides of the transition~\cite{kjall_many-body_2014,luitz_many-body_2015,pietracaprina_forward_2016,schiffer_many-body_2019,xu_many-body_2019,chanda_time_2020}. Only recently, a finite-size scaling analysis of the multifractal properties has found an asymmetric criticality for the multifractal dimension~\cite{mace_multifractal_2019}, thus making an interesting connection  with the Anderson problem on random graphs~\cite{abou-chacra_selfconsistent_1973,mirlin1994distribution, monthus2008anderson, biroli2012difference,de_luca_anderson_2014, tikhonov_anderson_2016, tikhonov2016fractality, PhysRevB.96.201114, garcia-mata_scaling_2017, biroli2018delocalization, kravtsov2018non, tikhonov2019statistics, garcia-mata_two_2020,tarzia_many-body_2020} {where such unusual scaling properties were first found \cite{garcia-mata_scaling_2017,garcia-mata_two_2020}}.
Nevertheless, most numerical studies violate the Harris bound~\cite{harris_effect_1974,chandran_finite_2015}, and a fully consistent finite-size scaling theory is still missing~\footnote{Note that the very applicability of this bound has been criticized by C. Monthus in Ref.~\cite{monthus_many-body-localization_2016}.}.

In such a puzzling context, several progresses have been made to build an analytical theory, able to describe the ergodic-MBL transition~\cite{vosk_many-body_2013,huse_phenomenology_2014,pekker_hilbert-glass_2014,vasseur_quantum_2015,altman_universal_2015,vosk_theory_2015,potter_universal_2015,dumitrescu_scaling_2017,schiro_toy_2020}. The most successful description, based on the so-called "avalanche" scenario~\cite{thiery_many-body_2018,thiery_microscopically_2017}, proposes a phenomenological renormalization group (RG) treatment, working in the MBL regime where large insulating blocks compete with small ergodic inclusions. In this framework, recent works~{\cite{goremykina_analytically_2019, dumitrescu_kosterlitz-thouless_2019,morningstar_renormalization-group_2019}} found a RG flow of the Kosterlitz-Thouless (KT) form. The MBL phase is described as a line of fixed points with a vanishing density of ergodic inclusions and a finite typical localization length $\zeta$, which controls the spatial extension of the $l$-bits~\cite{serbyn_local_2013,rademaker_many-body_2017,imbrie_local_2017}. When the delocalization transition is approached from the MBL side, the typical localization length $\zeta$ grows {and reach a \textit{finite}} critical {value} $\zeta_c=\left(\ln 2\right)^{-1}$ at the transition point $h_c$, with a singular behavior~\footnote{See \cite{garcia-mata_two_2020} for a similar behavior at the Anderson transition on random graphs.}  
\be
{\zeta}^{-1}-\zeta_{c}^{-1}\propto {\sqrt{h-h_c}}.
\label{eq:KT}
\ee
As argued in Ref.~\cite{dumitrescu_kosterlitz-thouless_2019}, a diverging length scale $\sim \exp(c/\sqrt{h-h_c})$ should control finite-size effects, thus explaining some limitations in the numerics.

In this Letter, we show that one can overcome such finite size constraints by measuring spin polarization, a simple local observable of the $S=1/2$ Hamiltonian Eq.~\eqref{eq:H}. Our main finding is that the extreme statistics of the polarizations gives a direct access to the typical MBL localization length $\zeta$, which shows a very good  agreement with Eq.~\eqref{eq:KT}. We numerically find finite-size effects of KT type, controlled by a correlation length $\sim \exp(c/(h-h_c)^{\nu_\text{loc}})$, with $\nu_\text{loc} \simeq 0.5$.
Building on Refs.~\cite{garcia-mata_scaling_2017,mace_multifractal_2019,garcia-mata_two_2020}, a careful scaling analysis of our ED data reveals that they are best described by a volumic scaling variable ${\cal N}/\Lambda$ in the delocalized regime, where $\cal N$ is the Hilbert-space size and $\Lambda$ a non-ergodicty volume, while a logarithmic scaling variable $(\ln L)/\lambda$ dictates the behavior in the MBL regime, with $\lambda^{-1}\propto \zeta^{-1}-\zeta_{c}^{-1}$ following the square root KT singularity Eq.~\eqref{eq:KT}.\\

\begin{figure}[b!]
\includegraphics[angle=0,clip=true,width=\columnwidth]{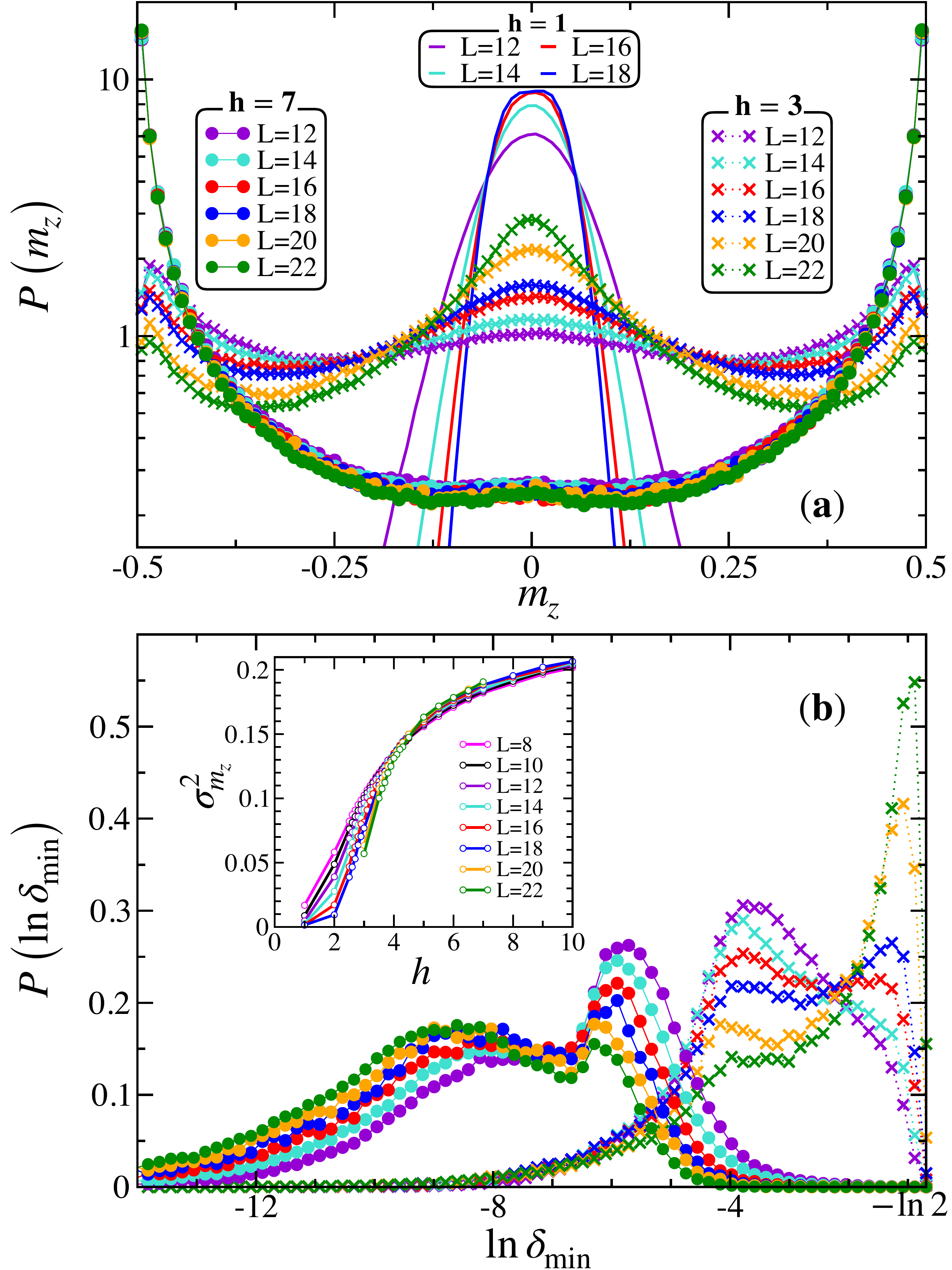}
\caption{Histograms of (a) the local magnetizations $m_z=\langle S_i^z\rangle$ displayed for $h=1,\,3,\,7$, and (b) of the maximally polarized sites Eq.~\eqref{eq:min}. Inset: variance $\sigma_{m_z}^{2}$ of the local magnetization plotted as a function of $h$ for various lengths $L$. Shift-invert ED results for infinite temperature eigenstates of Eq.~\eqref{eq:H}, performed over thousands of independent random samples. }
\label{fig:histo}
\end{figure}

\noindent{\bf Distribution of local polarizations and extreme value statistics---} 
Let us start the discussion by looking at the local polarizations $m_z=\langle S^z\rangle$, computed for individual infinite temperature eigenstates using ED simulations for the Hamiltonian Eq.~\eqref{eq:H}. As already noticed in Refs.~\cite{khemani_obtaining_2015,lim_many-body_2016,dupont_many-body_2019,hopjan_many-body_2019}, the histograms $P(m_z)$ display distinct features across the two regimes, as illustrated in Fig.~\ref{fig:histo}~(a) for 3 representative values of the disorder strength ($h=1,\,3,\,7$). At weak disorder, we expect from the eigenstate thermalization hypothesis (ETH)~\cite{deutsch_quantum_1991,srednicki_chaos_1994,dalessio_quantum_2016} to observe Gaussian distributions, peaked around $m_z=0$ and shrinking with increasing system size, as clearly visible for $h=1$.
At $h=3$, a more complex form emerges with deviations from gaussianity~\cite{luitz_anomalous_2016,luitz_ergodic_2017}: strongly polarized ($m_z \simeq \pm 1/2$) sites appear, but their density shrinks down with system size (as evidenced by the decrease of magnetization variance, see inset).
At strong disorder $h=7$, the density of strongly polarized sites no longer shrinks down: ETH is violated and the distribution is U-shaped, almost free of finite-size effects. 

To quantify this effect, we introduce the deviation from perfect polarization $\delta =\frac{1}{2}-\left|m_z\right|$.
While $\delta >0$ for finite systems, $\delta$ becomes arbitrarily small for large disorder, as evidenced in Fig.~\ref{fig:histo} (a) at $h=7$ where one observes~\cite{SM}
\be
P\left(\delta\right)\propto \delta^{-1+\frac{1}{\gamma}}\quad (\delta\to 0),
\label{eq:Pdeltam}
\ee
with an exponent $\gamma\ge 0$ related to extreme value statistics~\cite{majumdar_extreme_2020}, as we discuss now.
In each finite sample, we define for the most polarized site 
\be
\delta_{\rm min}=\frac{1}{2}-\max_{1\le i\le L}\left|\langle S_i^z\rangle\right|,
\label{eq:min}\ee
whose distributions are shown at $h=3$ and $h=7$ in Fig.~\ref{fig:histo} (b).
There, we explicitly report two distinct trends with increasing system sizes $L$.
In the ETH regime ($h=3$), some weight is transferred with increasing $L$ towards small $-\ln\delta_{\rm min}$, with a peak developing at $-\ln 2$. The opposite is observed in the MBL regime ($h=7$), where weight is moved towards large values of $-\ln\delta_{\rm min}$.
In both cases we qualitatively spot that rare events of the competing phase (rare insulating bottenecks for the delocalized phase {\it{vs.}} thermal bubbles in the MBL regime) become less and less relevant upon increasing $L$.\\

Let us now turn to a more quantitative analysis of the extreme polarization finite size scaling.
At strong disorder, assuming that the magnetizations along the chain are independently drawn from the distribution $P(m_z)$ of Fig.~\ref{fig:histo} (a), the deviation from perfect polarization verifies $\int_{0}^{\delta_{\rm min}}P(x){\rm d}x\sim {1}/{L}$~\cite{majumdar_extreme_2020}, which for a power-law tail Eq.~\eqref{eq:Pdeltam}, yields
\be
\delta_{\rm min}(L)\sim L^{-\gamma}.
\label{eq:extreme}
\ee
Such a scaling has dramatic consequences since for any $\gamma\neq 0$, we expect $\delta_{\rm min}(L)\to 0$ when $L\to \infty$, meaning spin freezing, and therefore chain breaking at the thermodynamic limit.
\vskip .15cm

\begin{figure}[t!]
\includegraphics[angle=0,clip=true,width=\columnwidth]{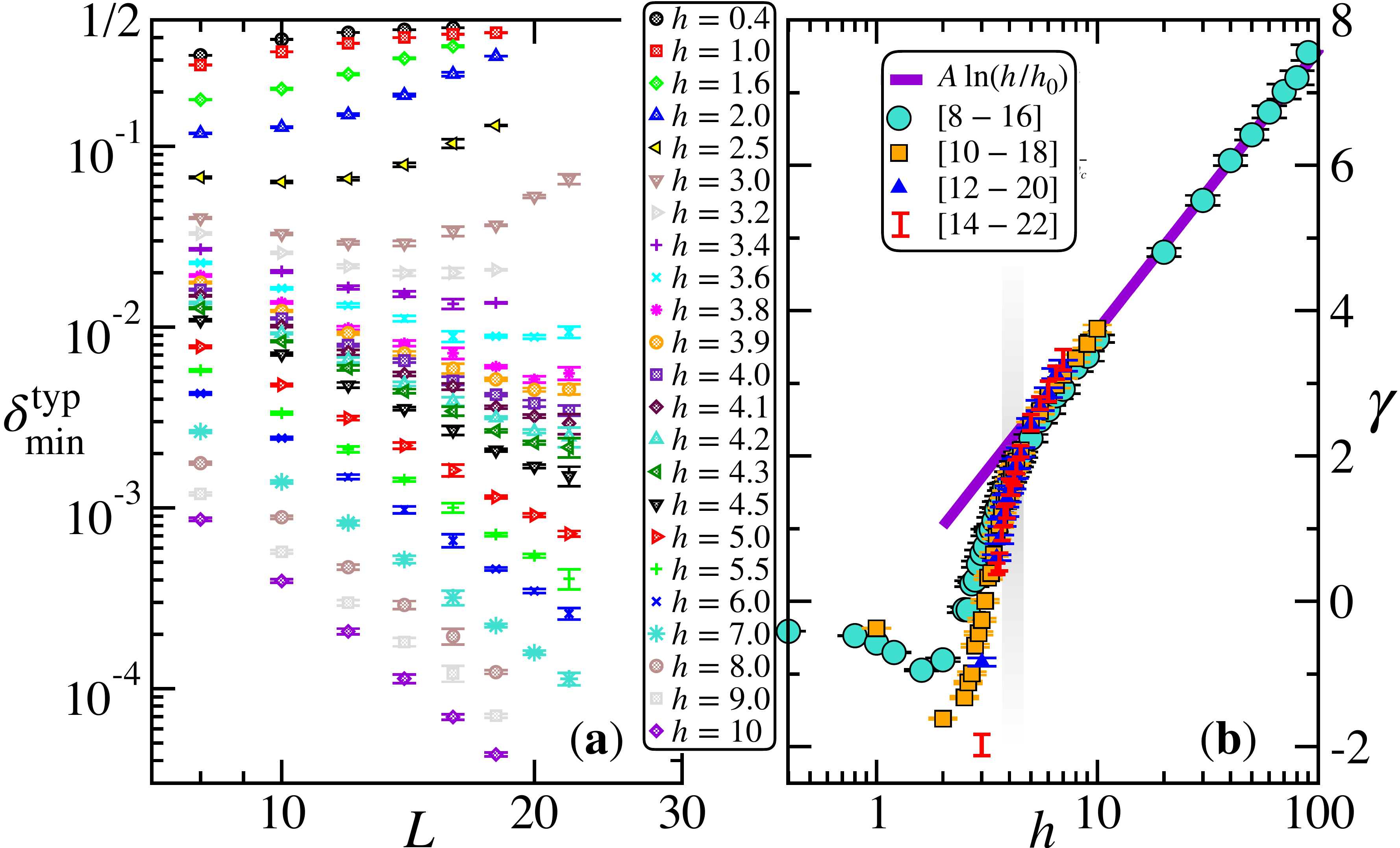}
\caption{(a) Typical deviation $\delta_{\rm min}^{\rm typ}$ Eq.~\eqref{eq:min}, plotted against system size $L$ for a wide regime of disorder $h$. Log-log scale reveals the power-law decay Eq.~\eqref{eq:extreme} at large enough $h$, while $\delta_{\rm min}^{\rm typ}\to 1/2$ at weak disorder. (b) Freezing exponent $\gamma$ governing the decay. Various symbols stand for 4 different fitting windows including 5 points in the range $[L_{\rm min},\,L_{\rm max}]$. At large disorder, the exponent grows logarithmically as $\gamma=A\ln (h/h_0)$ with fitting parameters $A=1.67(5)$, $h_0\approx 1$. The gray shaded area shows the critical region.}
\label{fig:delta_min}
\end{figure}

\noindent{\bf Extreme polarizations: numerical results---} 
In order to numerically check the power-law scaling Eq.~\eqref{eq:extreme}, and probe the chain breaking mechanism at the microscopic level, the most polarized site is recorded for each finite-length sample. Typical values are computed from shift-invert ED simulations at infinite temperature, 
and disorder averaging is performed over a large number of realizations, at least $10^3$ samples. Results as a function of chain lengths $L$ are shown in Fig.~\ref{fig:delta_min} (a) for a wide range of disorder strengths. 
As expected from the extreme value argument Eq.~\eqref{eq:extreme}, at strong disorder we observe a power-law decay with $L$, indicating a chain breaking at the thermodynamic limit. In contrast, for weak disorder $\delta_{\rm min}$ does not vanish with $L$, but instead tends to $1/2$, showing "healing" in a way similar to Kane-Fisher physics~\cite{kane_transport_1992,lemarie_kane-fisher_2019}. 
Such radically different behaviors correspond to the two phases of the model, as we argue below.

%

\begin{figure*}[ht!]
\includegraphics[angle=0,clip=true,width=2.05\columnwidth]{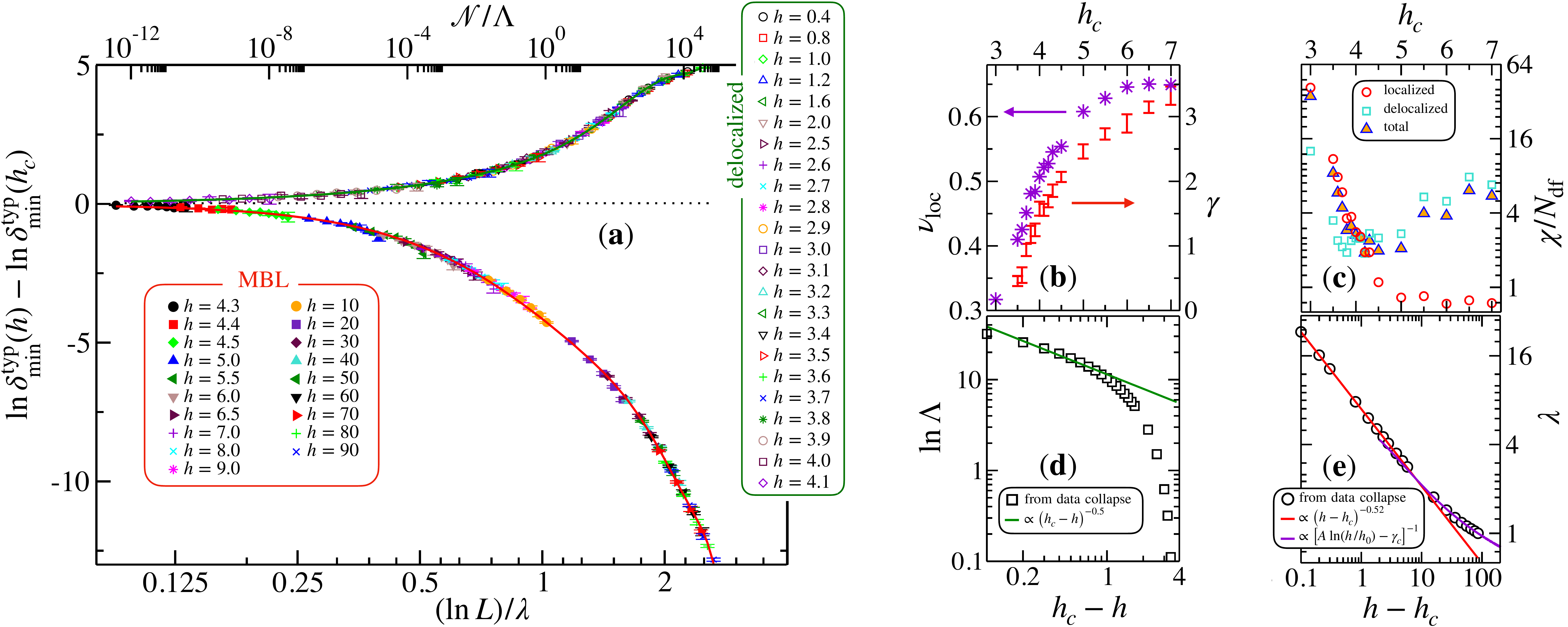}
\caption{(a) Scaling plots for both delocalized (top) and MBL (bottom) regimes following Eq.~\eqref{eq:scalings} with $h_c=4.2$. Green and red curves show the scaling functions $f$ and $g$ of Eq.~\eqref{eq:scalings} obtained from Taylor expansions close to $h_c$ fitted to the data~\cite{SM, PhysRevLett.82.382, PhysRevB.84.134209} with two distinct (disorder-dependent) scaling parameters: $\Lambda$ and $\lambda$ for the two phases. (d) The non-ergodicity volume $\Lambda$ diverges exponentially at criticality with an exponent {compatible with} $\nu_{\rm d}= 0.5$ (green line). (e) The MBL scaling parameter {$\lambda$} diverges at criticality as a power-law (red line) with an exponent $\nu_{\rm loc}=0.52(3)$, while there is a logarithmic dependence at strong disorder (violet line) showing perfect agreement with Fig.~\ref{fig:delta_min}~(b) using $A=1.67$, $h_0=1$, and $\gamma_c=1.7$. Panels (b) and (c) display {the outcome of such scaling procedures obtained for different choices of $h_c$}. In (b) the MBL exponent $\nu_{\rm loc}$ smoothly increases with $h_c$, {while in (c) the total chi-squared statistic for the best fit $\chi$ divided by the number of degrees of freedom $N_\text{df}$ (see text and \cite{SM}) displays an ${\cal{O}}(1)$ minimum} at $h_c=4.2$, where $\nu=0.52$. The freezing exponent $\gamma$ (red symbols) from Fig.~\ref{fig:delta_min} (b) is also shown in panel (b), and we get $\gamma_c=1.7(1)$.}
\label{fig:scaling}
\end{figure*}

In Fig.~\ref{fig:delta_min} (b), we show the $h$-dependence of the freezing exponent $\gamma$ obtained from power-law fits to the form Eq.~\eqref{eq:extreme} over 4 different fitting windows.
Finite-size scaling towards perfect polarization is governed by a disorder-dependent freezing exponent $\gamma$, which shows a logarithmic divergence at large $h$, a behavior explained below by analytical arguments. Conversely, at weak disorder we observe a finite-size (possibly non-monotonous, see also Ref.~\cite{PhysRevB.82.125455} for similar effects) crossover, and ultimately $\gamma\to 0$ as expected from the  ETH. Before making a quantitative scaling analysis of  ED data (Fig.~\ref{fig:scaling}), we first provide an analytical understanding of the chain breaking mechanism.
\vskip .15cm

\noindent{\bf Analytical derivation at large disorder---} 
Using the Jordan-Wigner transformation, we can rewrite Eq.~\eqref{eq:H} as interacting spinless fermions in a random potential
\be
{\cal{H}}=\sum_i\Bigl[\frac{1}{2}\left(c_{i}^{\dagger}c_{i+1}^{\vphantom{\dagger}}+c_{i+1}^{\dagger}c_{i}^{\vphantom{\dagger}}\right)-h_i n_i\Bigr]+\sum_i n_{i}n_{i+1}.
\label{eq:Hf}
\ee
The first sum, describing free fermions, can be diagonalized as ${\cal{H}}_0=\sum_{k=1}^{L}{\cal{E}}_k b_k^{\dagger} b_{k}^{\vphantom{\dagger}}$ with new fermionic operators $b_k=\sum_{i=1}^{L}\phi^k_ic_i$. The single-particle orbitals $\phi^k_i$ are exponentially localized for any $h \neq 0$, and the interacting Hamiltonian rewrites in this "Anderson basis" as 4 terms:
\bea
\label{eq:Hf}
{\cal{H}}&=&\sum_{k}\left({\cal{E}}_k+{\cal V}_{k}^{(1)}\right) n_k+\sum_{k\neq l}{\cal V}_{k,l}^{(2)}\,n_k n_l\\
&+&\sum_{k\neq l\neq m}{\cal V}_{k,l,m}^{(3)}\,n_k b_l^{\dagger} b_{m}^{\vphantom{\dagger}}+\sum_{k\neq l\neq m\neq n}{\cal V}_{k,l,m,n}^{(4)}\,b_k^{\dagger} b_l^{\dagger} b_{m}^{\vphantom{\dagger}}b_{n}^{\vphantom{\dagger}}\nonumber.
\eea
In the strong disorder limit $h\gg 1$, the second line can be neglected~\cite{SM}.
The number operators $n_k$ then form a complete basis of local operators commuting with $\mathcal{H}$ and among themselves: in other words, the Hamiltonian is written in $l$-bit form~\cite{serbyn_local_2013,huse_phenomenology_2014,chandran_constructing_2015,monthus_many-body_2016,rademaker_explicit_2016,imbrie_local_2017}, and the $n_k$ are good approximation of the $l$-bits in the strong disorder limit~\cite{de_tomasi_efficiently_2019}.
The fermion density {and thus the spin polarization} at real-space position $i$ is then simply the sum of contributions coming from occupied orbitals:
\begin{equation}
	\langle S_i^z \rangle + 1/2 = \langle n_i \rangle = \sum_{k~\text{occupied}} \left|\phi_i^k\right|^2.
	\label{eq:sum}
\end{equation}

Building on previous ideas~\cite{dupont_eigenstate_2019}, we are now ready to understand the spin freezing mechanism.
If we approximate the Anderson orbitals by simple exponential functions $|\phi^k_i|^2 \propto \exp\left(-\frac{|i-i_0^k|}{\zeta_0}\right)$, where $i_0^k$ is the localization center of orbital $k$, and $\zeta_0 \sim(\ln h)^{-1}$ the non-interacting localization length at strong disorder~\cite{dupont_eigenstate_2019}, we expect the maximal density $\langle n_i \rangle_{\rm max}$ to occur in the middle of the longest region of $\ell_\mathrm{max}$ consecutive sites that are occupied by an orbital. At half-filling, a configuration with $\ell$ consecutive occupied sites appears with probability $ 2^{-\ell}$, which for a finite chain of length $L\gg 1$ yields  $\ell_\mathrm{max}\approx \ln L/\ln 2$.

The most polarized site $i_{0}$ corresponds to the site with maximal $n_{i_{0}}\approx 1$ (resp. minimal $n_{i_{0}}\approx 0$) fermionic density, yielding $\langle S_{i_0}^{z}\rangle\approx +1/2$ (resp. $\langle S_{i_0}^{z}\rangle\approx -1/2$). As a result, the sum  of exponentials Eq.~\eqref{eq:sum}  gives $\delta_{\rm min} \sim \exp(-\ell_\mathrm{max}/(2\zeta_0))$, which naturally leads to the power-law decay Eq.~\eqref{eq:extreme} with a freezing exponent
\be
    \gamma(\zeta_0)\approx\frac{1}{2\zeta_0 \ln 2}.
    \label{eq:gamma_xi}
\ee
At large disorder, numerical data perfectly agree with a logarithmic growth $\gamma(h)\propto \ln h$ (panel (b) of Fig.~\ref{fig:delta_min}), which validates our analytical description of the freezing mechanism.
Within this description, $1/\gamma$ is identified with the localization length of $l$-bits deep in the bulk of the largest non-thermal region. Being far away from rare thermal inclusions, the inverse freezing exponent provides an estimate of the \emph{typical} $l$-bit extension, i.e. the \emph{typical} localization length $\zeta$: $1/\gamma \sim \zeta$.
As we further elaborate below, this has decisive consequences for our understanding of the critical behavior. Already in Fig.~\ref{fig:delta_min}~(b), when the  transition is approached from the MBL side, a singular behavior develops for $\gamma(h)$ {close to $h_c$}, in agreement with Eq.~\eqref{eq:KT}, followed by a jump to zero in the ergodic phase. We now provide an explicit description of this critical behavior.
\vskip .15cm

\noindent{\bf{Scaling analysis: KT behavior---}} We have performed a very careful analysis of our ED data for $\delta_{\rm min}^{\rm typ}(L,h)$ in order to address the transition between ergodic and MBL phases. The best data scalings~\cite{SM}, shown in Fig.~\ref{fig:scaling}~(a), are obtained using two distinct scaling functions:
\be
\ln\left[\frac{\delta_{\rm min}^{\rm typ}(L,h)}{\delta_{\rm min}^{\rm typ}(L,h_c)}\right]=\begin{cases}
    f({\cal N}/\Lambda)&  \text{if $h<h_c$},\\
    g\left(\frac{\ln L}{\lambda}\right)&  \text{if $h>h_c$}.
\end{cases}
\label{eq:scalings}
\ee
In the delocalized phase ($h<h_c$), we obtain a volumic scaling, in agreement with our recent multifractality analysis~\cite{mace_multifractal_2019} (see also \cite{garcia-mata_scaling_2017} for the Anderson transition on random graphs), and with ETH predictions \cite{deutsch_quantum_1991}, given that the density of state scales with Hilbert space size.
The scaling variable is the ratio between the Hilbert space size ${\cal N}\approx 2^L/\sqrt{L}$ and a disorder-dependent non-ergodicity volume $\Lambda$, which diverges exponentially at criticality: $\ln \Lambda\sim (h_c-h)^{-\nu_{\rm d}}$ with $\nu_{\rm d}\approx 0.5$, see Fig.~\ref{fig:scaling}~(d). The way the freezing exponent $\gamma\to 0$ in the ergodic regime (see Fig.~\ref{fig:delta_min}) is also controlled by ${\cal N}/\Lambda$.

Here {the fit of the scaling function to the data is} equally good for $h_c=3.8$ and $h_c=4.2$, as visible in Fig.~\ref{fig:scaling}~(c) where {the chi-squared statistic divided by the number of degrees of freedom of the fit} $\chi^2/N_{\rm df}$ {is minimum}~\cite{SM}.

In contrast, the MBL regime ($h>h_c$) is best described by the function $g\left(\frac{\ln L}{\lambda}\right)$ displayed in Fig.~\ref{fig:scaling}~(a).
This is a direct consequence (see \cite{mace_multifractal_2019, garcia-mata_two_2020, SM}) of the power-law decay Eq.~\eqref{eq:extreme}, observed at criticality and in the MBL phase, which yields an MBL scaling function $g\propto(\ln L)/\lambda$ for large $\ln L \gg \lambda $. As an outcome, the scale $\lambda$ is directly related to the freezing exponent, and {therefore} with the typical localization length, such that 
\be
{1}/{\lambda}\propto \gamma_c-\gamma\propto \zeta^{-1}_{c}-\zeta^{-1}_{\vphantom c}.
\ee
The larger $h_c \ge 4.5$, the better the goodness of fit for the MBL scaling, see Fig.~\ref{fig:scaling}~(c). Indeed, by definition the logarithmic scaling perfectly describes data with algebraic behavior such as $\delta_{\rm min}^{\rm typ}$ in the MBL phase.
To correctly estimate $h_c$, we therefore need to sum the goodness of fit from both ergodic and MBL regimes, see Fig.~\ref{fig:scaling} (c).
Then a bootstrap analysis of the data~\cite{SM} gives a critical disorder strength $h_c=4.2(5)$.
This quite large relative error of $\approx 10\%$ reflects the difficulties inherent to this ergodic-MBL transition.

However, this very number is not decisive for our theoretical 
understanding of the critical point~\cite{SM}.
As shown for $h_c = 4.2$ in Fig.~\ref{fig:scaling}~(e), we get
the following singularity for the inverse typical localization length:
\be
\gamma_c-\gamma\propto (h-h_c)^{\nu_{\rm loc}},
\label{eq:gamma_c}
\ee
with $\nu_{\rm loc}=0.52(3)$ after a bootstrap analysis, and 
$\gamma_c=1.7(1)$ (see panel (b) of Fig.~\ref{fig:scaling}).
This behavior quantitatively agrees with the recently proposed~\cite{dumitrescu_kosterlitz-thouless_2019,morningstar_renormalization-group_2019} KT mechanism, see Eq.~\eqref{eq:KT}. Moreover, the scaling variable $({\ln L})/\lambda$ implies that finite size effects in real-space are formally  controlled by an exponentially diverging length scale $\exp (\lambda)\sim \exp(h-h_c)^{-\nu_{\rm loc}}$, thus confirming a KT scenario, compatible with the Harris bound. This behavior contrasts with the multifractal behavior discussed in Ref.~\cite{mace_multifractal_2019}, controlled by a different length scale which probes the most delocalized regions (the thermal bubbles).

Before concluding, one should note that the scaling that we propose is unusual and it is not excluded that a different finite-size scaling may also be valid in the vicinity of the transition. Nevertheless, our assumptions have been quantitatively checked against all our numerical data.

\vskip .15cm
\noindent{\bf{Consequences and discussion---}}
Our key-result is that the extreme value statistics in the MBL regime gives a direct access to the typical localization length $\zeta$, and therefore the typical $l$-bits extension of the MBL states.
Maximally polarized sites correspond to entanglement bottlenecks: nearly frozen spins are almost disentangled from the rest of the system, with an entanglement entropy cutting such sites given by $S = -\delta \ln \delta$  to leading order~\cite{qie}.
Consequently, $\delta$ controls the leading eigenvalues of the entanglement spectrum, whose observed power-law distribution~\cite{samanta_extremal_2020,pal_probing_2020} is also governed by the freezing exponent $\gamma$.

Our analysis further indicates that the entire MBL regime, including the critical point, witnesses a chain breaking mechanism at the thermodynamic limit, with finite size effects controlled by a power-law behavior Eq.~\eqref{eq:extreme}.
The power-law exponent, related to the typical localization length $\gamma\propto \zeta^{-1}$,  diverges logarithmicaly at strong disorder, and displays a jump at the transition, with a singular critical behavior Eq.~\eqref{eq:gamma_c} in perfect agreement with a KT mechanism Eq.~\eqref{eq:KT}.

Besides probing the transition universality class, spin freezing  has deep consequences for MBL physics.
Firstly, it provides a simple picture accounting for the absence of thermalization at the thermodynamic limit.
Another important aspect concerns the recently discussed Hilbert-space fragmentation~\cite{roy_exact_2019,roy_percolation_2019,tomasi_dynamics_2019,pietracaprina_hilbert_2019} of the MBL regime, which here is expected to naturally emerge from spin freezing upon increasing system sizes. In addition, our results are very encouraging for the development of a perturbative decimation method~\cite{igloi_strong_2005} discarding the strongly polarized sites. Coupled to exact methods this could provide quantitative results at strong disorder for system sizes much larger than currently accessible, and thus potentially useful beyond one dimension.

{\it Note Added---} During the completion of this work, we became aware of a numerical study~\cite{Vidmar2020} which gives indications in favor of a KT transition, based on standard observables (e.g. entanglement entropy, level statstics). Note also the recent RG work~\cite{PhysRevB.102.125134} which suggests that the MBL transition is in a universality class that is slightly different from KT.

\begin{acknowledgments} This work is supported by the French National Research Agency (ANR) under projects THERMOLOC ANR-16-CE30-0023-02, {MANYLOK ANR-18-CE30-0017 and GLADYS ANR-19-CE30-0013}. We gratefully acknowledge F. Alet and M. Dupont for collaborations on related works. Our numerical calculations strongly benefited from the HPC resources  provided by 
CALMIP (Grants No. 2018- P0677 and No. 2019-P0677) and GENCI (Grant No. 2018- A0030500225). 
\end{acknowledgments}
%

\newpage
$~$
\newpage
\setcounter{figure}{0}
\setcounter{equation}{0}
\setcounter{table}{0}

\renewcommand{\thefigure}{S\arabic{figure}}
\renewcommand{\theequation}{S\arabic{equation}}
\renewcommand{\thesection}{S\arabic{section}}
\renewcommand{\thesubsection}{s\arabic{subsection}}
\renewcommand{\thesubsubsection}{s\arabic{subsubsection}}

\onecolumngrid

\begin{center}

{\large \bf Supplemental Material for "Chain breaking and Kosterlitz-Thouless scaling at the many-body localization transition"}
\vspace{0.6cm}

\end{center}

\twocolumngrid
{\bf{In this supplemental material we provide additional informations regarding:
1.~The power-law divergence of the distributions $P(\frac{1}{2}-\left|m_z\right|)$, in particular finite-size effects.
2.~The analytical derivation in connection with free-fermions.
3.~{The finite-size scaling analysis of the MBL transition}.}}
\section{Power-law divergence of the distributions: finite-size effects}
\begin{figure}[b!]
\includegraphics[angle=0,clip=true,width=1.035\columnwidth]{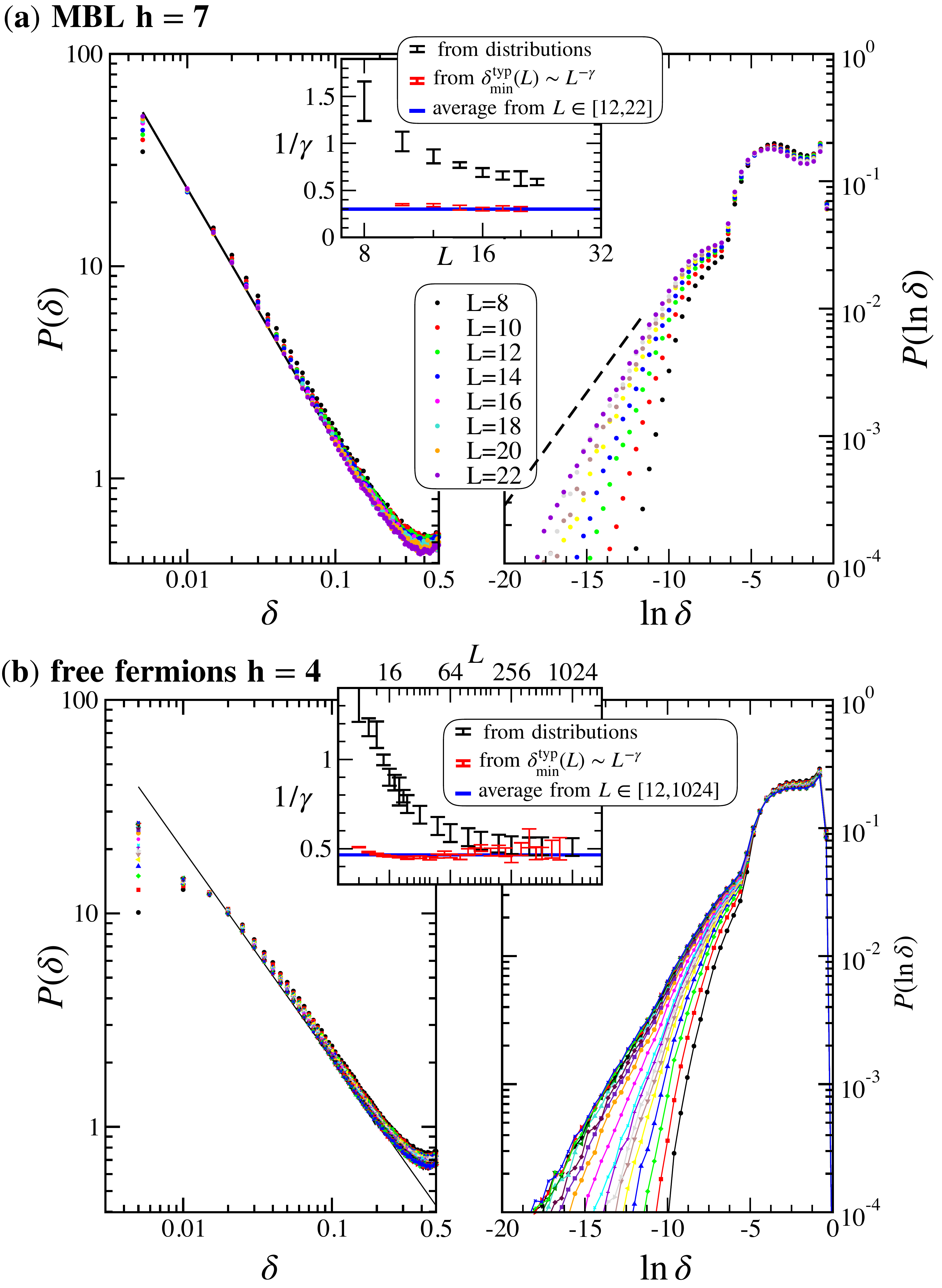}
\caption{Distributions of the deviations from perfect polarization $\delta=1/2-|m_z|$ for (a) MBL at $h=7$, and (b) non-interacting XX model Eq.~\eqref{eq:XX} at $h=4$. Log-log plot on the left panels focus on the non-universal power-law behavior for $P(\delta)$, diverging with an exponent $\approx 1$ (black line), while the right panels shows $P(\ln \delta)$, see Eq.~\eqref{eq:Pdeltam_exp}, which focuses on the universal tail. For the MBL case (a), the slope $1/\gamma$ slowly decreases towards its asymptotic value (thick blue line), as visible in the inset (black symbols) where $1/\gamma$ estimates from the extreme value scaling are also shown for comparison: either from 3-point fits (red symbols) of $\delta_{\rm min}^{\rm typ}(L)$ to the form Eq.~\eqref{eq:extreme}, or from a global fit (thick blue line) including all sizes $L\ge 12$. The slow convergence is better visible for the non-interacting problem (b) with the same system sizes as for the MBL case, together with the additional sizes $L=24,\,32,\,48,\,64,\,96,\,128,\,192,\,256,\,384,\,512,\,1024$. }
\label{fig:dist}
\end{figure}

In the MBL regime, the deviation form  perfect polarization $\delta =\frac{1}{2}-\left|m_z\right|$ displays power-law tails in the distributions, according to
$P\left(\delta\right)\propto \delta^{-1+\frac{1}{\gamma}}\quad (\delta\to 0)$,
which corresponds to an exponential tail for $\ln\delta$:
\be
P\left(\ln \delta \right)\propto \exp\left({\frac{\ln\delta}{\gamma}}\right).
\label{eq:Pdeltam_exp}
\ee
Histograms of ED data for $L=8,\,10,\,12,\,\ldots\,,22$ are shown in Fig.~\ref{fig:dist} (a) in the MBL regime ($h=7$). In the left panel, $P(\delta)$ is displayed in log-log scale, and a power-law behavior is visible, but does not correspond to the extreme value tail which we aim at exploring in this work. In order to better see such a tail, we show $P(\ln\delta)$ in the right panel where an exponential tail is clearly visible, thus targeting much smaller values of $\delta$, and thus the extreme values. However, the apparent prefactor $1/\gamma$ extracted from the exponential tail shows important finite-size effects, as reported in the  inset of Fig.~\ref{fig:dist} (a) where we have compared this estimate with the one extracted from the extreme value scaling 
\be
\delta_{\rm min}(L)\sim L^{-\gamma},
\label{eq:extreme}
\ee
which has almost no finite-size effects. 
Considering the limited  system sizes available to shift-invert ED techniques for the interacting problem, we have also explored this effect for the non-interacting case of many-body free-fermions, described by the random-field XX model
\be
{\cal{H}}_0=\sum_{i=1}^{L}\left(S_i^x S_{i+1}^{x}+S_i^y S_{i+1}^{y}-h_iS_i^z\right)
\label{eq:XX}
\ee
in the $S_{\rm tot}^{z}=0$ sector. Using standard free-fermion techniques, one can easily access to large systems, typically $L\sim 10^3$ without particular numerical effort. Results are shown in Fig.~\ref{fig:dist} (b) for the very same system sizes as for the MBL case, with the following additional lengths
$L=24,\,32,\,48,\,64,\,96,\,128,\,192,\,256,\,384,\,512,\,1024$.~Disorder strength $h=4$ ensures that non-interacting fermions have short localization lengths. Interestingly, we clearly observe a slow convergence of the tail exponent $1/\gamma$, as compared to the extreme value Eq.~\eqref{eq:extreme} which, like for the MBL case, shows almost no finite-size effects. Therefore, we clearly see that  focusing on the extreme value $\delta_{\rm min}(L)$ offers the most direct way to get the asymptotic value $\gamma$, as compared to the distribution tail.

\section{Analytical derivation in the Anderson basis}
\subsection{Fermionic representation of the spin problem}
The spin-$\frac{1}{2}$ XXZ Hamiltonian 
\bea
{\cal H}&=&\sum_{i=1}^{L}\Bigl[\frac{1}{2}\left(S_i^+ S_{i+1}^{-}+S_i^- S_{i+1}^{+}\right)-h_iS_i^z\nonumber\\
&+&\Delta S_i^z S_{i+1}^{z}\Bigr],
\label{eq:XXZ}
\eea
can be decomposed in two terms: ${\cal H}={\cal H}_0+V$. 

${\cal H}_0$ is the random-field XX model, Eq.~\eqref{eq:XX}, which corresponds to free-fermions (using the Jordan-Wigner transformation) in a random potential 
\be
{\cal H}_0=\sum_i\Bigl[\frac{1}{2}\left(c_{i}^{\dagger}c_{i+1}^{\vphantom{\dagger}}+c_{i+1}^{\dagger}c_{i}^{\vphantom{\dagger}}\right)-h_i n_i\Bigr].
\label{eq:H0}
\ee
The interaction terms read
\be
V=\Delta\sum_{i}^{L}n_i n_{i+1} +{\rm{C}},
\label{eq:V}
\ee
where C is an irrelevant constant. The non-interacting part Eq.~\eqref{eq:H0} is diagonalized by $b_k=\sum_{i=1}^{L}\phi^k_ic_i$ ($\phi_k$ being the single particle orbitals, Anderson localized for any finite disorder in 1D), yielding
$$
{\cal H}_{0}=\sum_{k=1}^{L}\epsilon_k b^{\dagger}_{k}b_{k}^{\vphantom{\dagger}}.$$
In this Anderson basis, the interaction term Eq.~\eqref{eq:V} is
\bea
V&=&\sum_{k,l,m,n}V_{k,l,m,n}b^{\dagger}_{k}b^{\vphantom\dagger}_{l}b_{m}^{{\dagger}}b_{n}^{\vphantom{\dagger}},\nonumber\\
{\rm{with~~}}V_{k,l,m,n}&=&\Delta\sum_{i=1}^{L}\phi^k_i\phi^l_i\phi^m_{i+1}\phi^n_{i+1},
\label{eq:Vanderson}
\eea
where we have assumed real orbitals.
As proposed in the main text, we decompose the interacting part in 4 terms:
\bea
V&=&\sum_k{\cal V}^{(1)}_{k}n_k\nonumber\\
&+&\sum_{k\neq l}{\cal V}^{(2)}_{k,l}n_k n_l\nonumber\\
&+&\sum_{k\neq l\neq m}{\cal V}^{(3)}_{k,l,m}n_kb^{\dagger}_{l}b^{\vphantom\dagger}_{m}\nonumber\\
&+&\sum_{k\neq l\neq m\neq n}{\cal V}^{(4)}_{k,l,m,n}b^{\dagger}_{k}b^{\vphantom\dagger}_{l}b^{\dagger}_{m}b^{\vphantom\dagger}_{n}.
\label{eq:all_terms}
\eea
The first two terms ${\cal V}^{(1,2)}$ are diagonal, and can be interpreted as a first approximation for the so-called $l$-bit Hamiltonian, while ${\cal V}^{(3,4)}$ are off-diagonal. 

\subsection{Strong disorder limit}
Below, we show that off-diagonal terms can be ignored at strong disorder. In such a limit, the Hamiltonian remains diagonal in the Anderson basis, i.e.
\be
{\cal H}_{\rm diag}=\sum_k\left(\epsilon_k+{\cal{V}}^{(1)}_{k}\right)n_k +\sum_{k,l}{\cal{V}}^{(2)}_{k,l}n_k n_l.
\ee
In this case, the maximum of the fermionic density can be readily obtained. We first use a simplified expression for the non-interacting orbitals, assuming the following exponential form 
\be
|\phi_i^k|^2=\tanh\left(\frac{1}{2\xi}\right)\exp\left(-\frac{|i-i_0^k|}{\xi}\right).
\label{eq:phik}
\ee
The particle density at a given site $i$ given by
\be
\langle n_i\rangle =\sum_{{\rm occupied~}k}|\phi^k_i|^2,
\ee
 is maximal in the middle of a region of $\ell_{\rm max}$ consecutive occupied orbitals. Indeed, for  $i\sim\ell_{\rm max}/2$ we have 
\be
n_\mathrm{max} \simeq \sum_{r=-\ell_{\rm max}/2}^{\ell_{\rm max}/2}\tanh\left(\frac{1}{2\xi}\right)\exp\left(-\frac{|r|}{\xi}\right),
\ee
where we have neglected the contribution of orbitals outside of the occupied region of size $\ell_{\rm max}$.
This yields for the deviation $\delta_{\rm min}=1-n_\mathrm{max}$ the following finite-size scaling
\bea
\delta_{\rm min}&\propto& L^{-\gamma}\nonumber\\
{\rm with~}\gamma&\approx&\frac{1}{2\xi \ln 2}\nonumber\\
&\approx&\frac{\ln h}{\ln 2}.
\label{eq:deltamin}
\eea
In the above expression we have used the fact that at strong disorder $\xi \approx (2\ln h)^{-1}$. Indeed, a simple perturbative expansion of any wavefunction away from its localization center yields amplitudes vanishing $\sim h^{-2r}$, where $r$ is the distance to the localization center.\\

We now provide simplified expressions for various terms in Eq.~\eqref{eq:all_terms} in the limit $h\gg 1$. In particular we justify why it is safe to ignore off-diagonal terms which vanish at large $h$.
\subsection{Diagonal terms}
In the large system size limit we have for the one-body term
\bea
{\cal V}^{(1)}_{k}&=&\Delta\frac{\tanh^2\left(\frac{1}{2\xi_k}\right)}{\sinh\left(\frac{1}{\xi_k}\right)}\nonumber\\
&\propto&\Delta\exp\left(-\frac{1}{\xi_k}\right) \propto \Delta/h^2\quad{\rm when~}{h\gg 1}.
\eea
The two-body contributions, assuming constant $\xi_k$ (this is justified at strong disorder where the distribution $P(\xi)$ is strongly peaked) reads
\bea
{\cal V}^{(2)}_{k,l}&=&\Delta\tanh^2\left(\frac{1}{2\xi}\right)\Bigl[r\exp\left(-\frac{r-1}{\xi}\right)+\frac{\exp\left(-\frac{r}{\xi}\right)}{\sinh\xi^{-1}}\Bigr]\nonumber\\
&\propto&\Delta\frac{r}{h^{2(r-1)}}\quad{\rm when~}{h\gg 1},
\eea
where $r=|i_0^k-i_0^l|\ge 1$ is the distance between two orbitals $k$ and $l$. Therefore the two-body interaction, which reads
\bea
\sum_{k\neq l}{\cal V}^{(2)}_{k,l}n_k n_l\approx \Delta\sum_k\Bigl(n_k n_{k+1}&+&\frac{2}{h^2}n_k n_{k+2}\\
&+&\frac{3}{h^4}n_k n_{k+3}+\cdots\Bigr),\nonumber
\eea
is clearly dominated by the nearest neighbor repulsion $\Delta$.
\subsection{Off-diagonal terms} 
\begin{figure}[t!]
\includegraphics[angle=0,clip=true,width=\columnwidth]{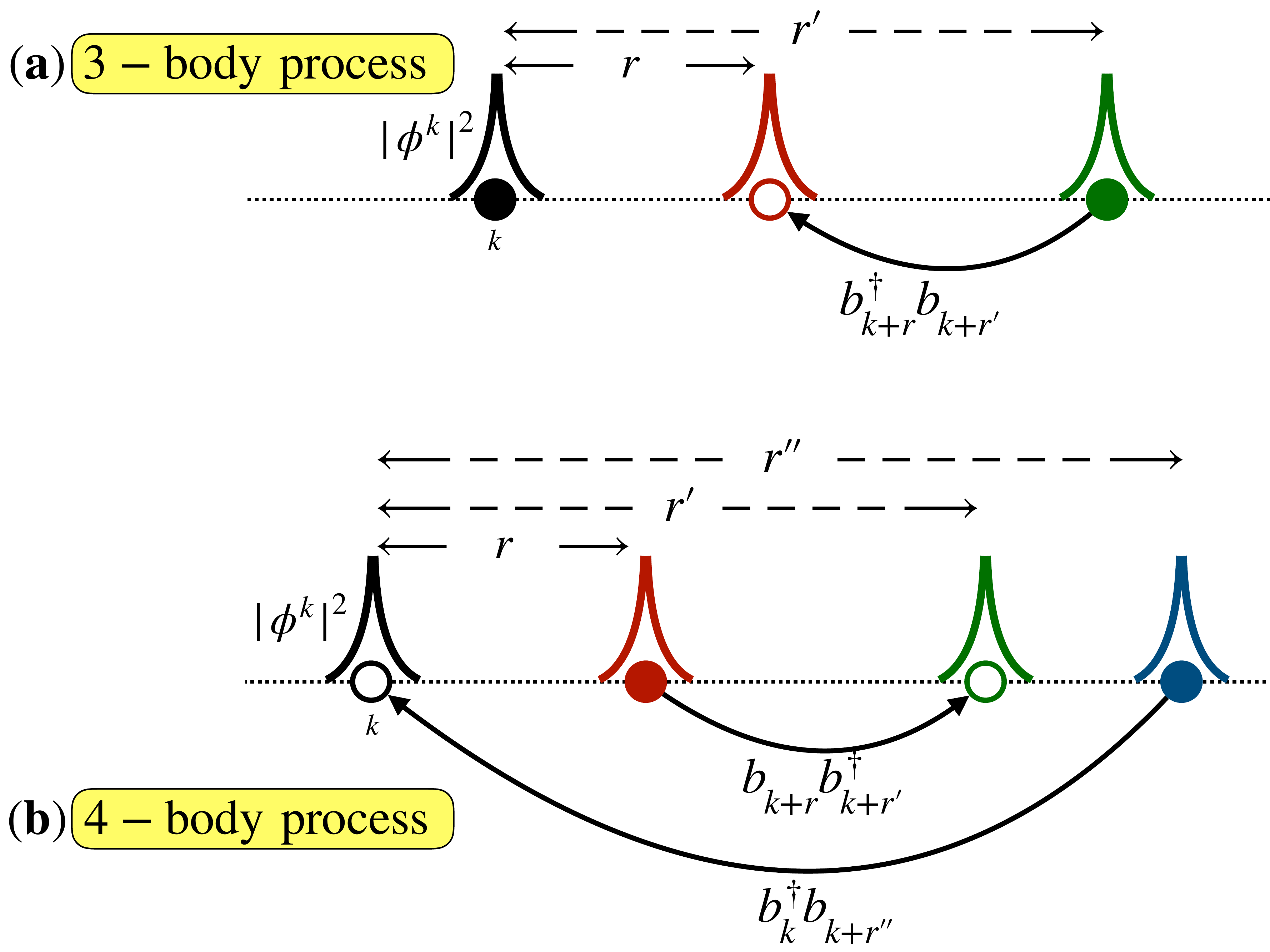}
\caption{Schematic picture of the 3- and 4-body processes: $3^{\rm rd}$ and $4^{\rm th}$ lines in Eq.~\eqref{eq:all_terms}.}
\label{fig:processes}
\end{figure}

We first discuss the three-body terms of the form
$${\cal V}^{(3)}_{k,l,m}n_kb^{\dagger}_{l}b^{\vphantom\dagger}_{m}.$$ From Eq.~\eqref{eq:Vanderson} we see that the amplitude can be randomly positive or negative. Nevertheless, one can easily anticipate that ${\cal V}^{(3)}_{k,l,m}$ will be dominated by maximally overlaping orbitals, more precisely by three nearest-neighbor orbitals $k,\,l,\,m$, such that $|i_0^k-i_0^l|=|i_0^l-i_0^m|=1$. In such a case
\be
\left|{\cal V}^{(3)}_{k,k+1,k+2}\right|\propto \frac{\Delta}{h}\quad{\rm when~}{h\gg 1},
\ee
where each orbital $k$ is labelled such that $\phi^k_i\sim \exp\left(-\frac{|i-k|}{\xi}\right)$, meaning that its localization center $i_0^k=k$.
In the more generic case, we have
\be
\left|{\cal V}^{(3)}_{k,k+r,k+r'}\right|\propto \frac{\Delta}{h^{r+r'-2}}\quad{\rm when~}{h\gg 1}.
\ee
A similar reasoning applies to the 4-body terms
$${\cal V}^{(4)}_{k,l,m,n}b^{\dagger}_{k}b^{\vphantom\dagger}_{l}b^{\dagger}_{m}b^{\vphantom\dagger}_{n},$$
with
\be
\left|{\cal V}^{(4)}_{k,k+r,k+r',k+r''}\right|\propto \frac{\Delta}{h^{r'+r''-r-2}}\quad{\rm if~}{h\gg 1}.
\ee
This contribution is dominated by situations where the 4 orbitals are nearest neighbors, i.e. when~$r=1,\,r'=2,\,r''=3$, thus giving 
\be
\left|{\cal V}^{(4)}_{\rm max}\right|\propto \frac{1}{h^2}.
\ee
When the relative distances between orbitals increases, the amplitudes ${\cal V}^{(4)}$ rapidly vanishes. Both 3-body and 4-body processes are illustrated in Fig.~\ref{fig:processes}.

\section{Finite-size scaling analysis of the MBL transition}

In this section, we detail our analysis by finite-size scaling of the polarization data. 

\subsection{Non-standard scaling approach}
The scaling theory of the MBL transition is subtle for a number of reasons: \begin{enumerate}
                                                                                                                            \item The finite-size scaling analysis of the MBL transition was first made as if it were a standard second-order phase transition, with a behavior of the type $y = y_c F (L / \xi)$ (where $y$ is the observable and $y_c=y(L,h_c)$) with $\xi$ a characteristic length diverging algebraically at the transition $\xi \sim | h-h_c | ^ {- \nu}$. Many works converged on a critical exponent $\nu \approx 1$~{\cite{kjall_many-body_2014,luitz_many-body_2015,pietracaprina_forward_2016,schiffer_many-body_2019,xu_many-body_2019,chanda_time_2020}}. This value of the critical exponent contradicts the Harris criterion~{\cite{harris_effect_1974,chandran_finite_2015}}.
                                                                                                                            
                                                                                                        \item Phenomenological RG studies~{\cite{vosk_theory_2015,potter_universal_2015,dumitrescu_scaling_2017, thiery_many-body_2018,thiery_microscopically_2017, goremykina_analytically_2019, dumitrescu_kosterlitz-thouless_2019,morningstar_renormalization-group_2019}} have suggested important finite-size effects with a critical exponent which has kept increasing up to very recently. Now, these approaches~{\cite{ goremykina_analytically_2019, dumitrescu_kosterlitz-thouless_2019,morningstar_renormalization-group_2019}} predict a length scale that diverges exponentially $\xi \sim e^{a | h-h_c | ^ {- 1/2}}$, with an RG flow that resembles that of the Kosterlitz-Thouless transition. It is well known that finite-size scaling analysis of KT type transitions is particularly delicate.
                                                                                                        \item An analogy exists between the MBL transition and the Anderson transition on random networks, a very rich problem, but nevertheless simpler, e.g. allowing analytical
                                                                                                        predictions~{\cite{abou-chacra_selfconsistent_1973,mirlin1994distribution, monthus2008anderson, biroli2012difference,de_luca_anderson_2014, tikhonov_anderson_2016, tikhonov2016fractality, PhysRevB.96.201114, 
                                                                                                        garcia-mata_scaling_2017, biroli2018delocalization, kravtsov2018non, tikhonov2019statistics, garcia-mata_two_2020}}. The scaling theory of the Anderson transition has only been clarified recently and is particularly subtle~{\cite{garcia-mata_scaling_2017,garcia-mata_two_2020}}. Thus according to the observables considered, there is a so-called `` linear '' scaling in $ y = y_c \, F (L / \xi) $ where L is the linear size of the random graph (ie its diameter), or a ``volumic'' scaling in $ y = y_c \,G (\mathcal N / \Lambda) $ where $\mathcal N$ denotes the volume of the graph (i.e. the number of sites) and $\Lambda$ is a characteristic volume diverging exponentially at the transition. Indeed, on the random graphs considered, the volume increases exponentially with the system size L and the linear or volumic scalings are not equivalent (contrary to the case of finite dimension). Different scalings have been found on both sides of the transition and two critical exponents exist in the localized phase: the average localization length diverges as $ \xi \sim (W-W_c)^{-1} $ ( where $W$ is the disorder strength) while the typical localization length follows a behavior: $ \zeta^{-1} = \zeta_c^{-1} + c \sqrt{W-W_c} $ (i.e. a critical exponent $1/2$) identical to the critical behavior recently predicted by phenomenological RGs for the MBL transition~{\cite{goremykina_analytically_2019, dumitrescu_kosterlitz-thouless_2019,morningstar_renormalization-group_2019}}. 
                                                                                                        
                                                                                                        \item Recently, the MBL transition has been analyzed in the Hilbert space where critical and MBL regions have been shown to host multifractal many-body states occupying a sub-volumic Hilbert space region $\mathcal N ^ {D_2} $, with $ D_2 <1 $ an $h$-dependent fractal dimension~\cite{mace_multifractal_2019}. This type of behavior is associated with a \emph{linear} scaling. In the ETH phase on the contrary, the scaling is found to be \emph{volumic}, indicating that the many-body states are ergodic in the Hilbert space at the thermodynamic limit.
                                                                                                        Supporting the random network analogy, the linear scaling of the MBL region is associated to a diverging length scale $\xi \sim (W - W_c)^{-0.7}$, while on the ETH side, the volumic scaling is associated to a diverging volume $ \Lambda \sim e^{a \vert h-h_c \vert ^ {-0.5}} $.

                                                                                                                                          \end{enumerate}
               
These observations invite us to consider non-standard scaling hypotheses: scaling in $\mathcal N / \Lambda$, in $L / \xi$ or even in $\ln L / \lambda$, which can be different on either side of the transition.
We will explain below the preferred scaling hypotheses and then describe the method we used to validate or not these hypotheses. Finally we will describe the results of these analyzes.

The phenomenological RG studies suggest very different critical and MBL behaviors for ``average'' observables, dominated by rare events (thermal bubbles), and for ``typical'' observables, controlled by the typical localization length $ \zeta $~{\cite{thiery_many-body_2018,thiery_microscopically_2017, goremykina_analytically_2019, dumitrescu_kosterlitz-thouless_2019,morningstar_renormalization-group_2019}}.
An idea to access the typical quantities is to consider observables that are the least affected by rare thermal bubbles.
Maximally polarized sites, being found in the bulk of localized regions, well separated from thermal inclusions, make up for a good candidate to probe typical properties. 
This is why we focused on $\delta_{\text{min}}^{\text{typ}}$.

\subsubsection{Scalings in the critical and MBL regions}
The power-law behavior $\delta_{\text{min}}^{\text{typ}} \sim L ^ {- \gamma} \sim \exp(-\gamma \ln L) $ observed at the largest sizes and predicted analytically at strong disorder suggests a scaling in $\ln L / \lambda$, since
{
\begin{equation}
	\ln \left[\frac{\delta_{\text{min}}^{\text{typ}}}{\delta_{\text{min}}^{\text{typ}}(h_c)} \right ] \sim -(\gamma - \gamma_c) \ln L \sim g\left( \frac{\ln L}{\lambda} \right),
\end{equation}}
with $g(X) \approx A0-A_1 X$ for large $\ln L\gg \lambda$.

Moreover, the observed power-law $\mathcal{N}^{D_2}$ for the eigenstates {multifractality on the Hilbert space \cite{mace_multifractal_2019} (see also \cite{garcia-mata_scaling_2017,garcia-mata_two_2020})}, which has been shown to be associated with the scaling $\ln \mathcal N/\xi$ indicates, with $L$ playing the role of $\mathcal N$, that the same hypothesis is also valid here.

{Note that we have observed a smooth dependence of the prefactor $A(h)$ on the algebraic law for $\delta_{\text{min}}^{\text{typ}} \approx A(h) L^{-\gamma(h)}$. Such dependence could be described by an irrelevant correction to the logarithmic scaling in the form $g(\ln L/\lambda)  + h(\ln L/\lambda)/\ln L  $ with $h(0)=0$ and $h(X) \approx CX$ for large $\ln L\gg \lambda$, $C$ a constant. Due to the limited range of variation of $L$, we did not take into account that correction.}

\subsubsection{Scalings in the delocalized region} In the ETH regime, canonical typicality {\cite{tasaki}} guarantees for local observables such as maximal polarization
$\delta_{\text{min}}^{\text{typ}}$ self-averaging with a variance $\propto \mathcal N^{-1/2}$~{\cite{PhysRevE.89.042112, PhysRevE.84.021130, PhysRevLett.108.110601, PhysRevE.92.020102, PhysRevB.93.134201, PhysRevE.99.032111}}.
We can therefore expect that this observable obeys a volumic scaling with $\mathcal N/\Lambda$, similarly to the participation entropies on the Hilbert space {\cite{mace_multifractal_2019}}.

\subsection{Controlled finite-size scaling approach}

To quantitatively test the compatibility of the numerical data with these different scaling hypotheses, we have adapted the controlled finite-size scaling approach {\cite{PhysRevLett.82.382, PhysRevB.84.134209}} as detailed below:
\begin{itemize}
 \item We first assume a value of $h_c$ and consider the data for the observable {$ 
\ln [\delta_{\text{min}}^{\text{typ}}/\delta_{\text{min}}^{\text{typ}}(h_c)]$} along with their uncertainties given by $\sqrt{\sigma^2 + \sigma_c^2}$, where $\sigma$ ($\sigma_c$, respectively) is the standard deviation of $\ln [\delta_{\text{min}}^{\text{typ}}]$ ($\ln[ \delta_{\text{min}}^{\text{typ}}(h_c)]$). We aim at asserting whether these data can be fitted by a function of the form:{
\begin{equation}
\ln \left [\frac{\delta_{\text{min}}^{\text{typ}}}{\delta_{\text{min}}^{\text{typ}}(h_c)}\right] = G(\rho \mathcal S^{1/\nu}) \; ,
\end{equation}}
where $\mathcal S$ can be either $\ln L$, $L$ or $\mathcal N$, $\rho \sim (h-h_c)$ close to $h_c$, and $\nu$ is a critical exponent to be determined by the fitting procedure. Importantly, we will fit the data for $h>h_c$ independently from the data for $h<h_c$, as they can follow different scalings. 
\item A controlled finite-size scaling analysis~{\cite{PhysRevLett.82.382, PhysRevB.84.134209}} consists in a Taylor development of the scaling function around $h=h_c$:
\begin{equation}\label{eq:scafunc}
G(\rho \mathcal S^{1/\nu}) = \sum_{j=0}^{n} a_j \rho^j \mathcal S^{j/\nu} \; ,  
\end{equation}
with 
\begin{equation}
 \rho (h) = (h-h_c) + \sum_{j=2}^{m} b_j (h-h_c)^j 
 \;.
\end{equation}

\begin{figure}[t!]
\includegraphics[width=\columnwidth]{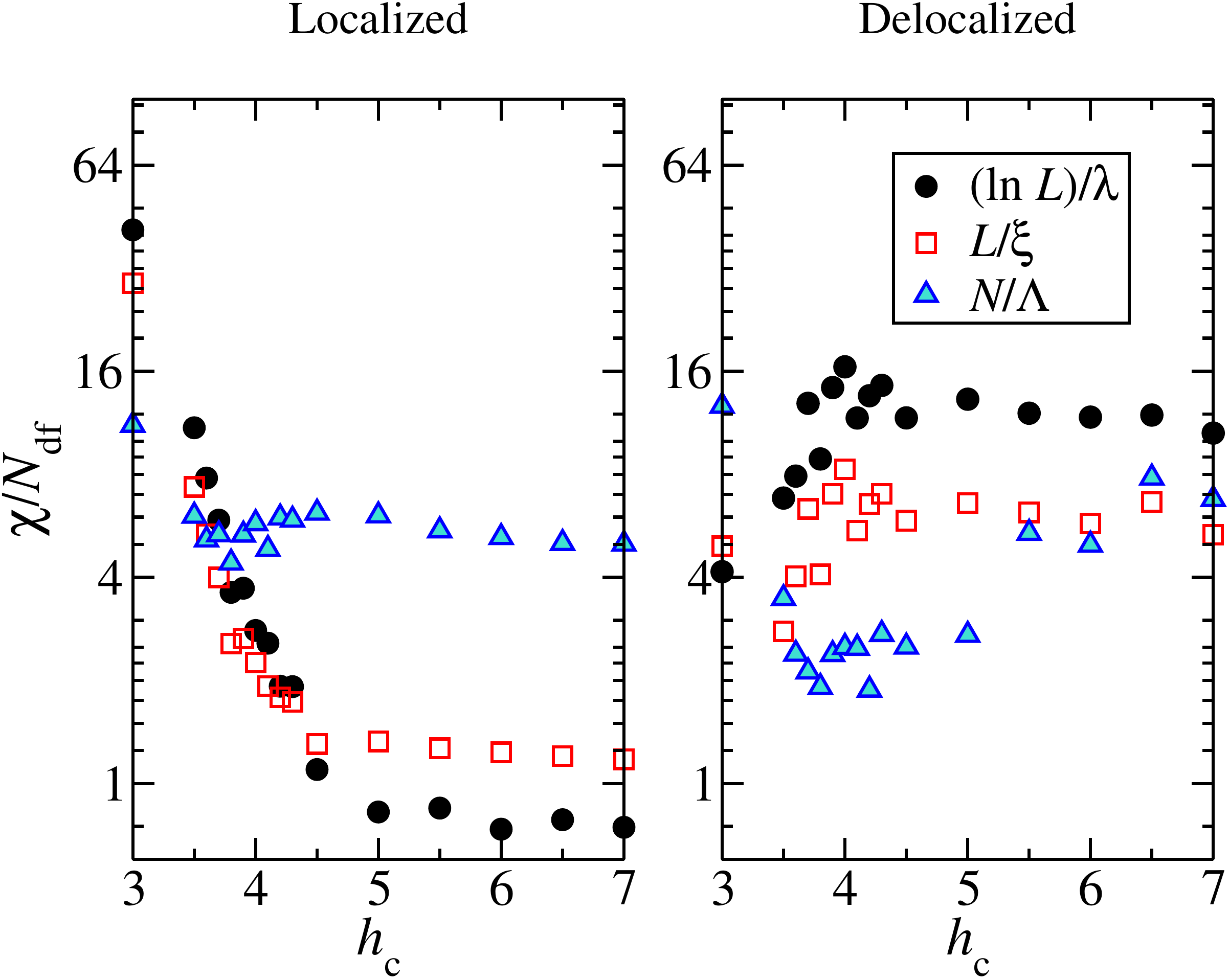}
\caption{{Quantitative estimate of the compatibility between the numerical data and the scaling assumptions. The localized ($h>h_c$) and delocalized ($h<h_c$) data have been fitted independently by Eq.~\eqref{eq:scafunc} with $\mathcal S=\ln L$, $L$ or $\mathcal N$. The value $\chi$ of the chi-squared statistic for the best fit divided by the number of degrees of freedom $N_\text{df}$ is plotted as a function $h_c$. $\chi/N_\text{df}$ should be of order one for an acceptable fit. Volumic scaling $\mathcal N/\Lambda$ prevails in the delocalized regime, while the data in the localized regime are compatible with a scaling as $\ln L/\lambda$. The best value of $h_c$, corresponding to the minimum of $\chi/N_\text{df}$, corresponds to $h_c =3.8$ and $h_c=4.2$ for the delocalized data, while the scaling of the localized data does not allow to determine $h_c$ (see text). If we combine the localized and delocalized data, the optimal $h_c$ is $h_c=4.2$ (see Fig.~3(c)).}}
\label{fig:chi2ovndf}
\end{figure}

The orders of expansion have been set to $n=5$ and $m=3$. The total number of parameters to be determined in the fit is $N_p=n+m+1$ (including $\nu$). We assessed the goodness of the fit by calculating the value $\chi$ of the chi-squared statistic for the best fit divided by the number of degrees of freedom $N_\text{df}=N_D-N_p$ where $N_D$ is the number of data, which should be of order one for an acceptable fit.  
\item We then tested systematically different values of $h_c$ and different scaling hypotheses, i.e. different choices $\mathcal S=\ln L$, $L$ or $\mathcal N$ for the localized $h>h_c$ and delocalized $h<h_c$ phases. The best value of $h_c$ corresponds to the minimum of $\chi/N_\text{df}$, taking into account only the localized or delocalized data, or all the data by considering $\chi^\text{tot}/N_\text{df}^\text{tot} \equiv (\chi^\text{loc}+\chi^\text{deloc})/(N_\text{df}^\text{loc}+N_\text{df}^\text{deloc})$.

\item To determine the uncertainties on $h_c$ and $\nu$, we have used a bootstrap procedure. From the data, we generated 100 \textit{synthetic} data sets by sampling randomly from normal distributions centered on the true data and with standard deviations given by the $\sigma$s. We then fitted these data sets and calculated the $\chi/N_\text{df}$. The best value of $h_c$ for each synthetic data set was determined and the uncertainty corresponds to the standard deviation of these $h_c$s. The fluctuations of $\nu$ within the data sets was also recorded.  
\end{itemize} 

\subsection{Results}

We present here additional results to the ones reported in the letter. The analysis by finite-size scaling has been done without excluding any size or value of the disorder. The sizes are between $ L = 8 $ and $ L = 22 $ and the disorder between $ h = 0.4 $ and $ h = 90 $.

The figure \ref{fig:chi2ovndf} represents $\chi/N_\text{df}$ as a function of $h_c$ in the localized ($h>h_c$) and delocalized ($h<h_c$) regimes. This quantitative test of the different scaling hypotheses confirm our expectation that volumic scaling $\mathcal N/\Lambda$ prevails in the delocalized regime, while the data in the localized regime are compatible with a scaling as $\ln L/\lambda$. 

It is not surprising to see this $\ln L/\lambda$ scaling in the localized phase favors large values of $h_c$, since it is by construction good for algebraic data, which is the case for large values of $h$.

On the delocalized regime, the volumic scaling has a optimum around $h_c =3.8$ and $h_c=4.2$, with a value of $\chi^\text{deloc}/N_\text{df}^\text{deloc}= 1.9$ which is systematically reduced if we exclude the lowest values of $ h $.

If we combine the localized and delocalized data, the optimal $h_c$ is $h_c=4.2$ with a $\chi^\text{tot}/N_\text{df}^\text{tot} = 1.9$ which is remarkably low given the considerable ranges of variation of $h$.

This scaling procedure assumes a power law divergence of the characteristic volume $\Lambda$ in the delocalized phase. This is done for practical reasons. We have not been able to implement an exponential divergence in this procedure, but we have added non-linear corrections for $\rho$ to allow for a more complex behavior than just a strict {power-law} divergence.
The best fit gives a very large value of the critical exponent $\nu_\text{d}\approx 8.$ which indicates that the divergence of $\Lambda$ close to $h_c$ is most probably an exponential divergence. Indeed, the figure 3(d) shows that the data for $\Lambda$ are compatible with $\Lambda \sim e^{a (h_c-h)^{-0.5}}$. Such a behavior has been observed for the scaling of the participation entropy in the Hilbert space {\cite{mace_multifractal_2019}}, and is known to control the Anderson transition on random graphs in the delocalized phase {\cite{garcia-mata_scaling_2017,garcia-mata_two_2020}}. Importantly, the volumic scaling implies that the whole delocalized phase obeys ETH at the thermodynamic limit.

The bootstrap procedure gives for the delocalized data ($h<h_c$) an average $\langle h_c\rangle =3.96$ and a standard deviation of $\sigma_{h_c}=0.22$. For the localized data, $\langle h_c\rangle =6.22$, $\sigma_{h_c}=0.64$ which confirms that the scaling as $\ln L/\lambda$ is not able determine the critical value of $h_c$. However, combining the localized and delocalized data gives $\langle h_c\rangle =4.46$, $\sigma_{h_c}=0.32$. We therefore combine the delocalized and total estimations to give the estimation $h_c=4.2(5)$.

On the localized phase, the divergence of $\lambda \sim (h-h_c)^{-\nu_\text{loc}}$ corresponds to values of the critical exponent $\nu_\text{loc}\approx 0.5$ which drifts slowly as $h_c$ increases (see figure 3(b)). The uncertainty on $h_c$ translates into a larger uncertainty on $\nu_\text{loc}$ than that of the fluctuations of $\nu_\text{loc}$ found from the bootstrap procedure. We find $\nu_\text{loc}=0.52(3)$.
Importantly, this behavior is compatible with the predictions from phenomenological RGs~{\cite{goremykina_analytically_2019, dumitrescu_kosterlitz-thouless_2019,morningstar_renormalization-group_2019}}, with a characteristic length scale controlling finite-size effects in the MBL phase diverging exponentially as $e^{b (h-h_c)^{-0.5}}$ close to the transition. Together with the algebraic critical behavior $\delta_{\text{min}}^{\text{typ}}(h_c) \sim L^{-\gamma_c}$ and the relation found between $\gamma$ and the typical localization length $\zeta$, Eq.(9) of the letter, this also confirms that the typical localization length reaches a finite value at the transition with a square-root singularity, as predicted by phenomenological RGs~{\cite{ thiery_many-body_2018,thiery_microscopically_2017,goremykina_analytically_2019, dumitrescu_kosterlitz-thouless_2019,morningstar_renormalization-group_2019}}.

\subsection{Non-parametric finite-size scaling}
\begin{figure}[ht!]
\includegraphics[angle=0,clip=true,width=1.0\columnwidth]{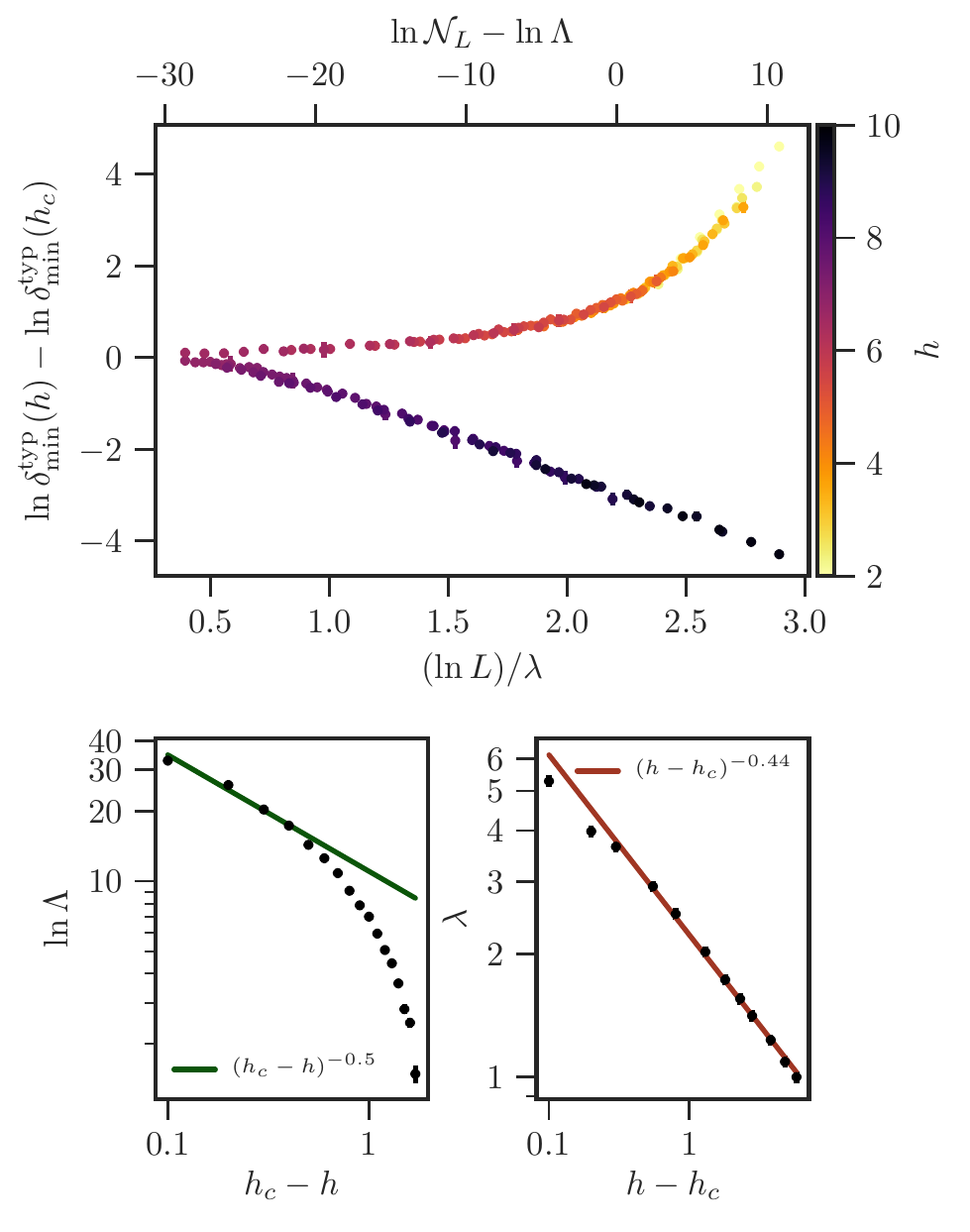}
\caption{Top figure: Scaling plots for both delocalized (top) and MBL (bottom) regimes with $h_c=4.2$. Best collapse is obtained via a non-parametric procedure (see text).
Bottom figures: $\Lambda$ (left) and $\lambda$ (right) obtained form the collapse. The non-ergodicity volume $\Lambda$ diverges exponentially with an exponent $\nu_{\rm d} \approx 0.5$ (green line), while the MBL scaling parameter {$\lambda$} diverges at criticality as a power-law (red line) with an exponent {$\nu_{\rm loc} \approx 0.44$}.}
\label{fig:scaling_nonparametric}
\end{figure}

In parallel with the controlled finite-size scaling approach whose results are presented above and in the main text, we have performed a non-parametric finite-{size} scaling.
This approach does not assume a parametrization for the scaling functions {$f$ and $g$ (see Eq.~(10) of the main text) and scaling parameters} $\Lambda(h)$ and $\lambda(h)$, but instead finds the best {collapse of the data} by optimizing an objective function which quantifies how close the data points are from being collapsed onto a unique curve.
The objective function we chose to optimize is Spearman's rank correlation coefficient, $R$, which is a measure of how close a cloud of points are from forming the graph of a monotonic function.
This approach did not allow us to quantitatively determine the best value of $ h_c $ nor the uncertainties on the critical exponent $\nu_\text{loc}$.
However, assuming $h_c = 4.2$, we obtain scaling results consistent with the parametric approach, as Fig.\ \ref{fig:scaling_nonparametric} shows.
In particular, we find a power-law divergence of the MBL scaling parameter with an exponent $\nu_{\rm loc} \approx 0.44$, close to the previously reported $\nu_{\rm loc} = 0.52$.
In the ETH region, we observe a behavior compatible with an exponentially divergent non-ergodicity volume with exponent $\nu_{\rm d} = 0.5$.

\end{document}